\documentclass[aps,prb,a4paper,10pt,twocolumn,showpacs,floatfix,bibnotes,superscriptaddress,preprintnumbers,longbibliography]{revtex4-2}
\usepackage{graphicx} 
\usepackage{dcolumn}  
\usepackage{bm}       

\usepackage{soul}
\usepackage{color}
\usepackage[colorlinks,bookmarks=false,citecolor=darkblue,linkcolor=red,urlcolor=blue]{hyperref}

\usepackage[charter,greekuppercase=italicized]{mathdesign}

\usepackage{microtype}
\usepackage{afterpage}
\graphicspath{{../}{./}{figures/}}

\definecolor{darkred}{rgb}{0.7,0.0,0.0}

\definecolor{darkblue}{rgb}{0,0.02,0.45}

\definecolor{darkgreen}{rgb}{0.02,0.45,0.0}

\definecolor{violet}{rgb}{0.8,0.2,0.6}

\begin{document}

\title{Frustration-induced quantum criticality in Ni-doped CePdAl as revealed by the $\mu$SR technique}

\author{I.\ Ishant}
\affiliation{\mbox{Department of Physics, Shiv Nadar Institution of Eminence, Gautam Buddha Nagar, Uttar Pradesh 201314, India}}

\author{T.\ Shiroka}
\email{toni.shiroka@psi.ch}
\affiliation{Laboratory for Muon-Spin Spectroscopy, Paul Scherrer Institut, CH-5232 Villigen PSI, Switzerland}
\affiliation{Laboratorium f\"ur Festk\"orperphysik, ETH Z\"urich, CH-8093 Z\"urich, Switzerland}

\author{O.\ Stockert}
\affiliation{Max Planck Institute for Chemical Physics of Solids, Dresden, Germany}

\author{V.\ Fritsch}
\affiliation{Experimental Physics VI, Center for Electronic Correlations and Magnetism, University of Augsburg, 86159 Augsburg, Germany}

\author{M.\ Majumder}
\email{mayukh.cu@gmail.com}
\affiliation{\mbox{Department of Physics, Shiv Nadar Institution of Eminence, Gautam Buddha Nagar, Uttar Pradesh 201314, India}}

\date{\today}

\begin{abstract}
In CePdAl, the 4$f$ moments of cerium arrange to form a geometrically
frustrated kagome lattice. Due to frustration in addition to Kondo- and RKKY- interactions, this metallic system shows
a long-range magnetic order (LRO) with a $T_{\rm N}$ of only 2.7\,K.
Upon Ni doping at the Pd sites, $T_{\rm N}$ is further suppressed, to
reach zero at a critical concentration $x_c \approx 0.15$. Here, by using
muon-spin relaxation and rotation ($\mu$SR), we investigate at a local
level CePd$_{1-x}$Ni$_x$Al for five different Ni-concentrations, both
above and below $x_c$. 
Similar to the parent CePdAl compound, for $x = 0.05$, we observe an
incommensurate LRO, which turns into a quasi-static magnetic order for
$x = 0.1$ and 0.14. More interestingly, away from $x_c$, for $x = 0.16$
and 0.18, we still observe a non-Fermi liquid regime, evidenced by a
power-law divergence of the longitudinal relaxation at low temperatures.
In this case, longitudinal field measurements exhibit a time-field scaling,
indicative of a cooperative spin dynamics that persists for $x > x_c$.
Furthermore, similar to the externally applied pressure, the chemical
pressure induced by Ni doping suppresses the region below $T^*$,
characterized by a spin-liquid like dynamical behavior. 
Our results suggest that the magnetic properties of CePdAl are similarly
affected by the hydrostatic- and the chemical pressure. We also
confirm that the unusual non-Fermi liquid regime (compared to conventional
quantum critical systems) is due to the presence of frustration that
persists up to the highest Ni concentrations.
\end{abstract}

\maketitle

\section{Introduction}

Heavy-fermion compounds with partially localized $d$- or $f$ magnetic 
moments have long been at the center of numerous theoretical- and
experimental studies~\cite{Qimiao2001}. The interest on these
strongly-correlated electron systems stems from the plethora of
phenomena they exhibit, such as quantum criticality~\cite{Brando2016,Coleman2005},
hidden order~\cite{Palstra1985}, charge-density wave~\cite{Hossain2005},
or the possibility to behave as superconductors~\cite{Steglich1979},
topological insulators~\cite{Jiang2010}, or topological Weyl
semimetals~\cite{Liu2021,Yang2021}.
In magnetic heavy-fermion compounds, two types of interactions dominate.
One of them is the Ruderman-Kittel-Kasuya-Yosida (RKKY) interaction,
where the conduction electrons mediate the interaction between the local
magnetic moments which, consequently, tend to form a long-range order.
If the coupling between the local moments and the conduction electrons
is $J$, the strength of the RKKY interaction varies as
$T_\mathrm{RKKY} \propto J^2$ ~\cite{Ruderman1954,Yosida1957}.
On the other hand, the Kondo interaction arises because of the
formation of many-body singlet states between the local moments
and the conduction electrons. Since it screens the local moments, it
tends to suppress the long-range magnetic ordering temperature. 
The strength of the Kondo interaction varies as
$T_\mathrm{K} \propto e^{(-1/J)}$~\cite{Stewart1984}. In 1977, Doniach
proposed a phase diagram which reflects the $J$ dependence of the two
interactions and indicates that, for small values of $J$, the RKKY
interaction dominates, while, for high values of $J$, the Kondo
interaction dominates~\cite{Doniach1977}. Because of the competition
between these two interactions, at a particular value of $J$, the
long-range magnetic ordering disappears. Thus, by utilizing a
non-thermal tuning parameter (such as pressure, magnetic field, or
doping), one can tune $J$ and continuously suppress the second-order
phase transition to $T=0$, to reach the so-called quantum
critical point (QCP). Quantum fluctuations at the QCP break down the
Fermi-liquid model, and a non-Fermi-liquid behavior~\cite{Schofield1999,Stewart2001}
is expected to show up in different physical properties, such as
magnetic susceptibility, heat capacity, spin-lattice relaxation rate, etc.

\begin{figure}
{\centering {\includegraphics[width=8cm]{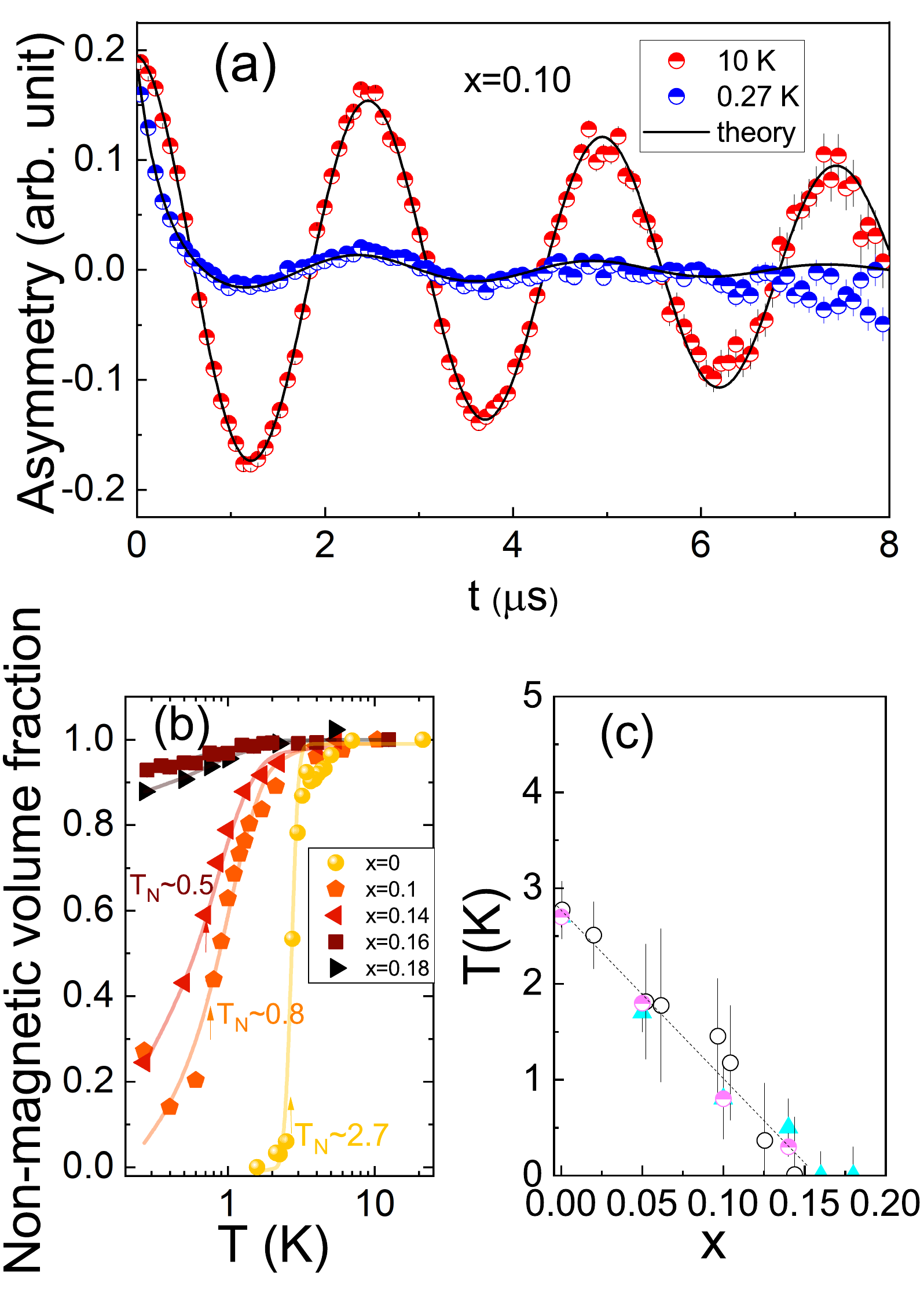}}\par} \caption{\label{fig:asym}(a) wTF-$\mu$SR asymmetry of $x = 0.1$ in a 3-mT applied field. (b) Nonmagnetic volume fraction vs.\ temperature estimated from wTF measurements for different Ni concentrations. Solid lines are fits using Eq.~(\ref{eq:sigmoidal_func}). (c) Temperature-concentration diagram of Ni-doped CePdAl resulting from wTF-$\mu$SR data. The dotted line is a linear fit. Triangles represent $T_{\rm N}$ from the present study, while heat-capacity data (open circles), and neutron-scattering results (half-filled circles) are taken from Ref.~\onlinecite{Fritsch2013,Huesges2017}, respectively.}
\end{figure}    

Recently, a ``global phase diagram'' has been proposed on top of the
Kondo- and RKKY interactions~\cite{Si2010}, incorporating also the
effects of quantum fluctuations (which increase with increasing frustration).
The inclusion of frustration, arising from local moments arranged into
a triangular, kagome, Shastry-Sutherland, or pyrochlore lattice, induces
different types of quantum criticality (local- and SDW-type)~\cite{Hertz1976,Millis1993}
and exotic phases of matter, e.g., quantum-spin liquids (QSL)~\cite{Vojta2008}.
A QSL is an entangled state where magnetic moments, despite being highly
correlated, do not break any symmetry, even down to $T = 0$.
Unlike in insulators, where QSL states are more common, in metallic systems
they are quite rare. Yet, in some frustrated metallic systems, such as
CeRhSn~\cite{Tokiwa2015}, Pr$_2$Ir$_2$O$_7$~\cite{Nakatsuji2006},
CeRh$_{1-x}$Pd$_x$Sn~\cite{Tripathi2022}, and CeIrSn~\cite{Shimura2021},
a QSL-type spin dynamics has been observed.
The study of metallic systems with magnetic moments 
lying in a frustrated lattice is comparatively less explored,
but nonetheless intriguing. For instance, 
in the absence of Dzyaloshinskii-Moriya interactions, skyrmion spin
textures have been observed in $4f$-based metallic systems with
a frustrated lattice~\cite{Okubo2012,Leonov2015}.

In this respect, CePdAl, where the Ce moments are arranged in a highly
frustrated kagome lattice~\cite{Fritsch2013}, is a very promising system
for exploring the above mentioned topics. CePdAl (with $T_\mathrm{N} \approx 2.7$\,K)
is also very susceptible to tuning. Thus, hydrostatic pressure can
suppress the LRO and a QCP emerges at a $P_c$ of about 0.9\,GPa.
More interestingly, an extended non-Fermi liquid regime with QSL-like
dynamic correlations has been observed up to about twice $P_c$. Such a
phase was claimed to be due to the presence of frustration~\cite{Majumder2022,Zhao2019}.
Thus, the presence of frustration not only stabilizes the QSL state,
but it also changes the generic features of QCP by producing an
extended NFL regime well beyond the critical point. As the lattice parameters were found to decrease with increasing Ni-doping \cite{Fritsch2013}, the resulting chemical pressure caused by Ni replacement at the Pd sites suppresses $T_{\rm N}$ to 0\,K at a critical concentration $x_c \approx 0.15$. In this case, the relevant
questions are: ``What is the nature of the critical fluctuations around QCP?
Is there any extended non-Fermi liquid regime/state also with
chemical pressure? What are the similarities or dissimilarities between
the hydrostatic and chemical pressure? Is frustration responsible for
the non-Fermi-liquid behavior, or the (chemical-doping induced) disorder
can account for it?'' To answer these questions, we employed a
well-known local-probe technique such as muon-spin relaxation/rotation
($\mu$SR). All samples were specimen of Czochralski-grown single crystals which were powderized and pressed into pellets for the measurements. The $\mu$SR experiments were performed on CePd$_{1-x}$Ni$_x$Al with five different Ni-concentrations $x$, both
above and below $x_c = 0.15$: 0.05, 0.1, 0.14, 0.16, and 0.18 on the Dolly spectrometer of the Swiss Muon Source (S$\mu$S) at the Paul Scherrer Institute, Switzerland. The base temperature for the Dolly spectrometer is 270 mK reached by using a $^3$He cryostat. Such an
extensive investigation allowed us not only to answer the above questions,
but also to construct a comprehensive phase diagram of CePdAl under Ni doping.

\section{Experimental results}

The NFL behavior in a metal arises from its low-lying quasi-particle
excitations, to which a low-frequency probe like $\mu$SR is quite
sensitive. Thus, extensive ($\mu$SR) measurements were performed to
elucidate the complex phase diagram of CePd$_{1-x}$Ni$_x$Al.
The details of the different $\mu$SR experiments, along with the results
obtained, are discussed in the following sections.

\subsection{Weak transverse field (wTF) measurements}
Weak transverse field (wTF) measurements allow us to extract the temperature dependence of the non-magnetic volume fraction from which the magnetic ordering temperature can be estimated. The terms weak and transverse indicate that the applied field is less than the internal field (and low enough, so that it does not deviate the muon beam) and that the applied magnetic field is transverse with respect to the direction of the muon spin, respectively. The wTF spectra were  collected in a 3-mT applied field at several temperatures and in samples with different Ni-concentrations. Figure~\ref{fig:asym}(a) depicts the wTF asymmetry of the $x = 0.1$ sample at 0.27 and 10\,K. The wTF asymmetries were fitted by a combination of an exponentially relaxing oscillatory component, representing the paramagnetic spins, and a fast exponentially relaxing non-oscillatory component, representing the spins taking part in the magnetic ordering [solid lines in Fig.~\ref{fig:asym}(a)]:
\begin{equation}
A(t) =  A_{0} [f_\mathrm{pm}\cos (\omega t + \phi_\mathrm{TF})e^{-\lambda_\mathrm{pm} t} + (1-f_\mathrm{pm})e^{-\lambda_\mathrm{ord} t}].
\end{equation}
Here, $A(t)$ is the time-dependent muon asymmetry, with $A_0$ the initial asymmetry. $f_\mathrm{pm}$, $\omega$, $\phi_\mathrm{TF}$, $\lambda_\mathrm{pm}$ are the fraction of spins, the muon Larmor frequency, the initial phase, and the relaxation rate in the paramagnetic state. Finally, $\lambda_\mathrm{ord}$ is the relaxation rate in the magnetically ordered phase.
At the highest temperature, the externally applied field dominates the
internal field ($H_\mathrm{ext} \gg H_\mathrm{int}$), thus giving rise
to asymmetry oscillations with the highest $f_\mathrm{pm}$, indicative
of a fully paramagnetic state. As $T$ decreases, the oscillation amplitude
$f_\mathrm{pm}$ decreases, indicating the onset of a spontaneous internal
field. At the same time, ($1-f_\mathrm{pm}$) increases as the temperature
decreases, indicating the enhanced magnetic contribution related to the
long-range magnetic order. As shown in Fig.~\ref{fig:asym}(b), the
temperature dependence of $f_\mathrm{pm}$ (the non-magnetic volume fraction)
was fitted with an empirical sigmoidal function:
\begin{equation}
\label{eq:sigmoidal_func}
    A(T) = A_a + \frac{A_b - A_a}{1 + \exp\frac{T-T_{\rm N}}{\Delta T}}
\end{equation}
Here, $T_{\rm N}$ is the ordering temperature and $\Delta T$ is the
transition width. $A_a$ and $A_b$ are the asymmetries above and below
$T_{\rm N}$, respectively. The suppression of $T_{\rm N}$ of CePdAl as
the Ni doping increases is shown in Fig.~\ref{fig:asym}(c). Our results
are consistent with those from magnetization- and heat-capacity measurements~\cite{Fritsch2013,{Huesges2017}}.
The linear suppression of $T_{\rm N}$ with Ni doping yields a
critical concentration $x_c \approx 0.15$.

\subsection{Zero-field measurements}

Zero-field (ZF) $\mu$SR measurements are sensitive to small magnetic
moments (as small as 0.001 $\mu_\mathrm{B}$) and, thus, are a powerful
tool to study weak internal fields and a possible magnetic order. The
time dependence of ZF-$\mu$SR asymmetry provides useful hints into
the nature of the long-range ordered state and the quantum critical
behavior of the electron spin dynamics close to the QCP.

\subsubsection{Magnetically ordered state (for $x < x_c$)}
\begin{figure}
{\centering {\includegraphics[width=8.5cm]{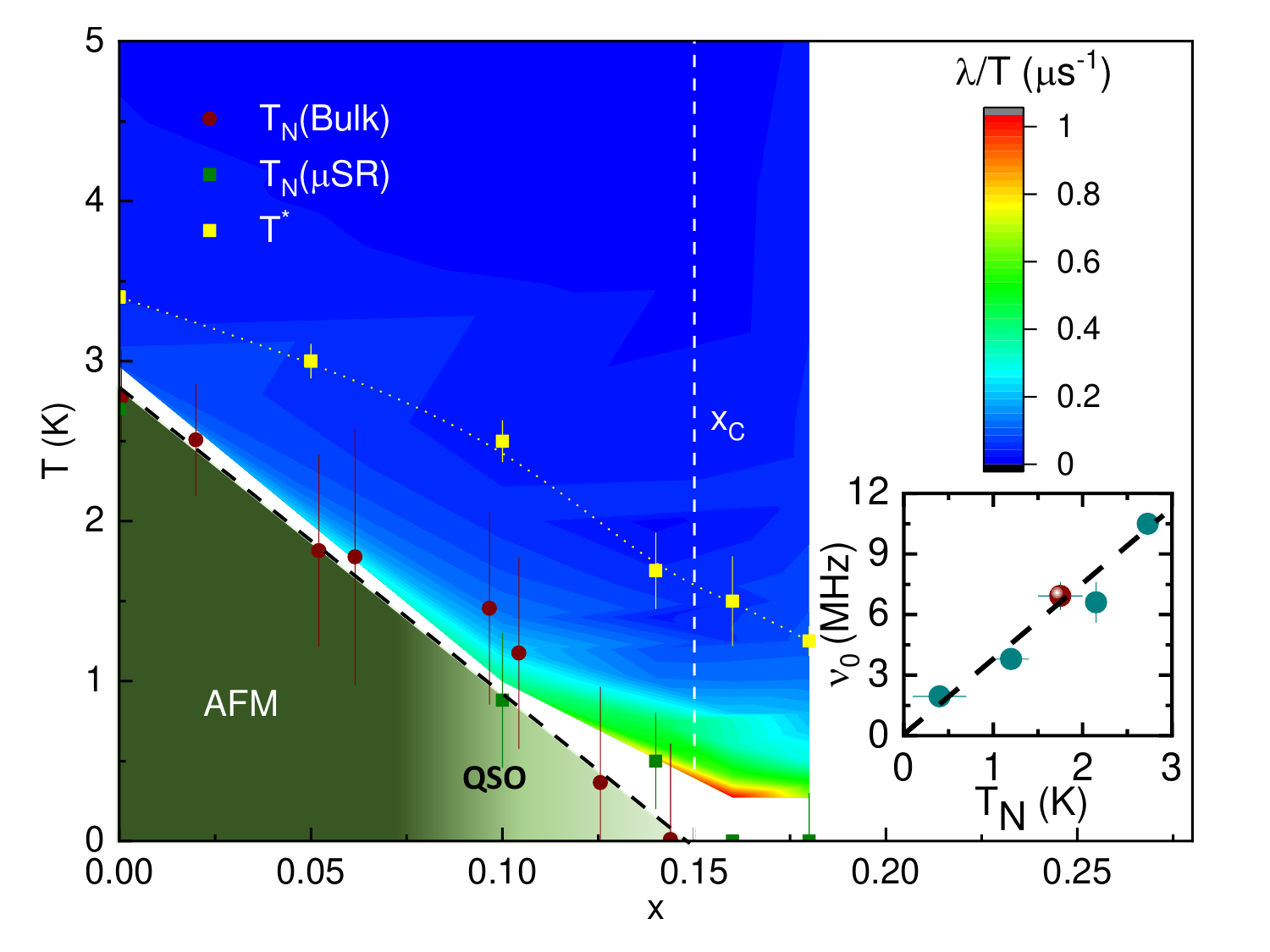}}\par} \caption{\label{fig:phase-diagram}Temperature-concentration magnetic phase diagram of Ni-doped CePdAl combined with a contour plot of the concentration and temperature-dependent relaxation rate $\lambda/T (x,T)$. The Neel temperatures $T_{\rm N}$ estimated from wTF-$\mu$SR (green squares, with error bars indicating the width of transition) agree well with the reported values from susceptibility measurements (filled circles). Yellow squares represent $T^*(x)$. QSO stands for quasi-static magnetic ordering, as discussed in the text. Inset shows $\nu_0$ vs.\ ordering temperature $T_{\rm N}$ (the dashed line is a linear fit). The brown- and the cyan circles correspond to the doping- and pressure evolution of $T_{\rm N}$ and $\nu_0$, respectively.}
\end{figure}

Figure~\ref{fig:zf_musr} shows the ZF-$\mu$SR asymmetry spectra of all
the samples ($x = 0.05$, 0.1, 0.14, 0.16, and 0.18), recorded at 0.27\,K.
For $x = 0.05$, the zero-field $\mu$SR asymmetry is very similar to that
of the CePdAl parent compound. In this case, the same function used
for CePdAl~\cite{Majumder2022} provides perfect fits:

\begin{figure}
{\centering {\includegraphics[width=8.5cm]{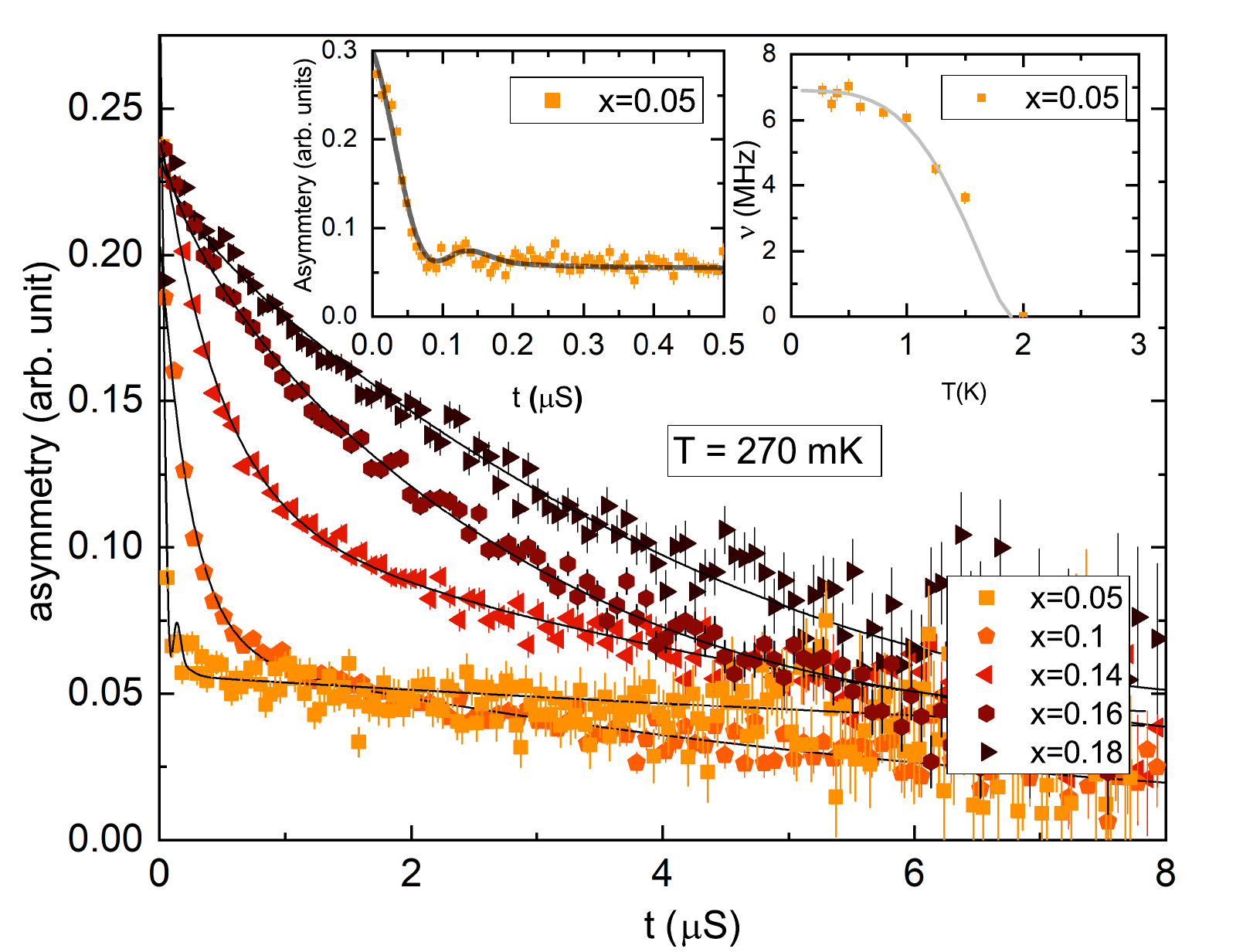}}\par} \caption{\label{fig:zf_musr}Zero-field $\mu$SR asymmetry for different Ni-concentrations. Solid lines are fits to the equations described in the text. Insets show the short-time $\mu$SR asymmetry (left) and $\nu$ vs.\ temperature (right) for the $x = 0.05$ case.}
\end{figure}
%
\begin{eqnarray}
A(t) & = & A_{0} \lbrace  f_\mathrm{sample}\left[ \frac{2}{3}j_{0}(2\pi\nu t + 
\phi)e^{\frac{-\sigma^2 t^2}{2}}+ \frac{1}{3}e^{\lambda_Lt}\right]  \nonumber \\ 
 & + & (1-f_\mathrm{sample})e^{-\lambda_\mathrm{bkg}t} \rbrace  
\end{eqnarray}
Here, $A_0$ is the initial asymmetry and $\sigma$ is the depolarization
rate caused by the distribution of internal fields. $\nu$ is the frequency
of the oscillating component and $\phi$ is its phase (here, kept at zero).
$\lambda_L$ is the longitudinal relaxation rate, $f_\mathrm{sample}$ is
the fraction of moments involved in the long-range order (LRO), and
$\lambda_\mathrm{bkg}$ the background relaxation contribution.
As shown in the inset of Fig.~\ref{fig:zf_musr}, below $T_\mathrm{N}$,
the order parameter can be mapped into the oscillation frequency, whose
temperature dependence for the $x = 0.05$ case is described by
$\nu(T) = \nu_0[1 - (T/T_\mathrm{N})^\delta]^\eta$, with $\delta = 3.5$
and $\eta = 2$. Here, $\nu_0$ is the local field at the muon site,
extra\-po\-la\-ted to $T = 0$. This is shown to vary linearly with the ordering
temperature $T_\mathrm{N}$ (see inset in Fig.~\ref{fig:phase-diagram}),
here, an implicit function of pressure~\cite{Majumder2022}. 
$T_\mathrm{N}$ is shown to correlate linearly also with the ordered
magnetic moment (estimated from neutron diffraction experiments~\cite{Huesges2017}),
thus strongly suggesting that, as expected, the local magnetic field at
the muon stopping site is proportional to the ordered moment. 

\begin{figure}
{\centering {\includegraphics[width=8.5cm]{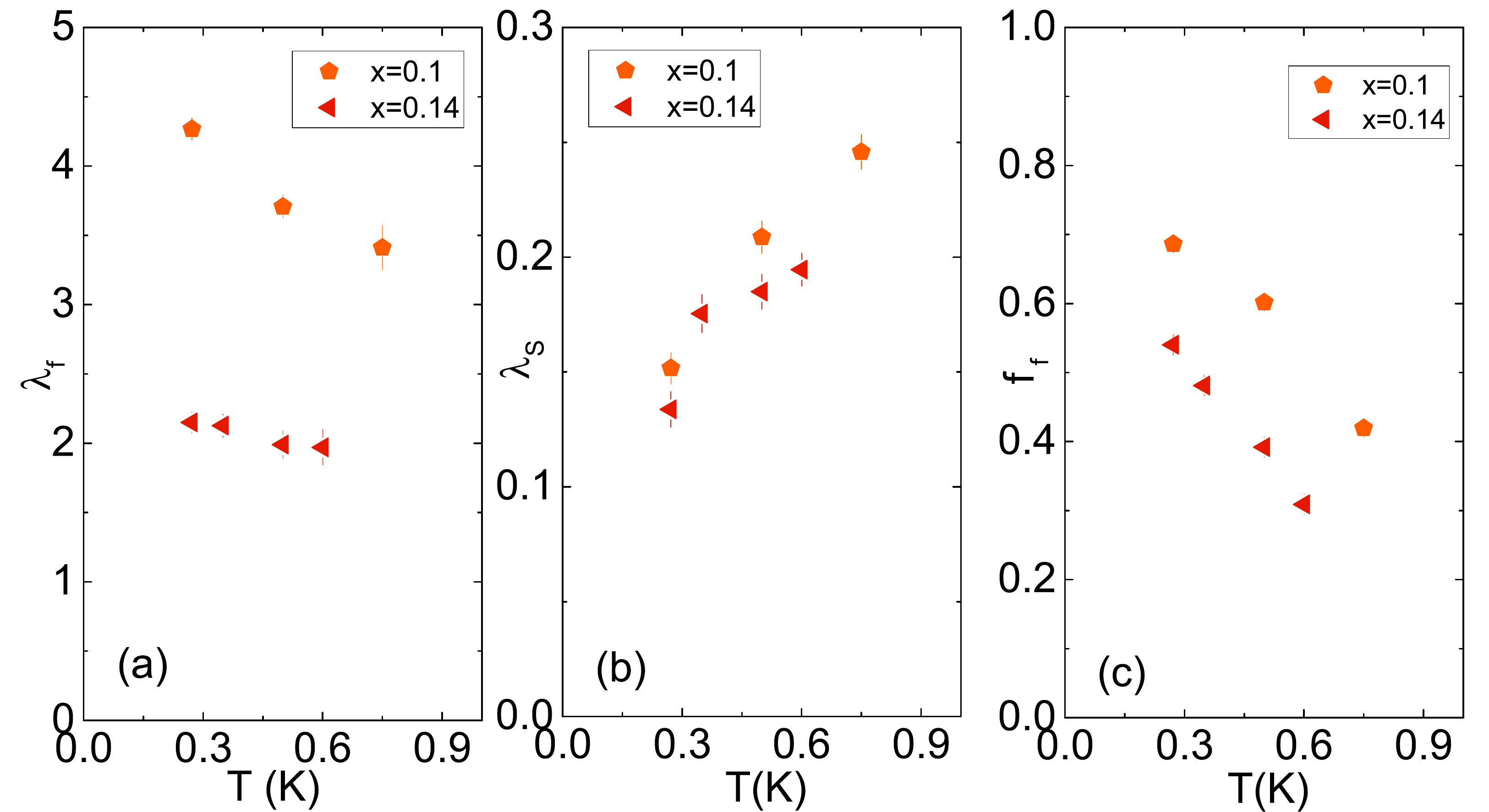}}\par} \caption{\label{fig:zf_pars_vs_T}Temperature dependence of $\lambda_{f}$ (a), $\lambda_{s}$ (b), and $f_f$ (c) for $x = 0.1$ and $x = 0.14$.}
\end{figure}

Interestingly, at higher Ni doping ($x = 0.1$ and 0.14), no spontaneous
muon spin precession is observed down to 0.27\,K, despite a magnetic 
ordering taking place at 0.8\,K and 0.5\,K, respectively (as determined 
from wTF-$\mu$SR data). Below these ordering temperatures, the zero-field
$\mu$SR asymmetry of the $x = 0.1$ and 0.14 samples is well fitted by the sum of
two exponential functions, one with a slow- and one with a fast 
relaxation rate: 
\begin{equation}
A(t) = A_0[f_{s}(e^{-\lambda_{s}T}) + f_{f}(e^{-\lambda_{f}T})]
\end{equation}
Here, $A_0$ is the initial asymmetry, fixed at the highest temperature.
$f_s$ and $f_f$ ($= 1-f_s$) are the fractions of the slow- and the fast
relaxation components, while $\lambda_s$ and $\lambda_f$ are the
respective muon-spin relaxation rates. The temperature dependence of
$\lambda_{f}$, $\lambda_{s}$ and $f_f$ is shown in Fig.~\ref{fig:zf_pars_vs_T}.
%
%
%
As the temperature is lowered, $\lambda_{f}$ and $f_f$ increase, while
$\lambda_{s}$ decreases. Since the $f_f$ fraction increases to
approximately 0.66 ($\sim 2/3$) and $\lambda_{f}$ shows an
order-parameter-like behavior, we identify the fast- and the slow
component with the transverse- and the longitudinal relaxation, respectively. 
Their behavior can be ascribed either to a quasi-static magnetic ordering,
as found also in other disordered systems, or to an incommensurate ordering,
the latter leading to a broad distribution of local magnetic fields at
the muon stopping site~\cite{Kenney2019,Kulbakov2021}.
\textcolor{black}{The incommensurate ordering, however, may be discarded for the present systems, as we observe clear oscillations in the asymmetry due to incommensurate
ordering at ambient- and under applied pressure~\cite{Majumder2022} which indicates that the local magnetic field distribution here is not broad enough to completely wipe-out the oscillation.} 
On the other hand, disorder originating from Ni doping is a more likely
explanation for the evolution of the long-range magnetic order (in 
$x = 0.05$) into a quasi-static magnetic order at higher Ni concentrations 
(see Fig.~\ref{fig:phase-diagram}).
The role of disorder is confirmed also by the increased broadening of
the magnetic Bragg peak at $Q_{AF}$ (0.5, 0, 0.35) with increasing Ni
content~\cite{Huesges2017}, a result at least partially attributed to disorder which
further corroborates our hypothesis. Note that, we attempted to fit the
ZF-$\mu$SR asymmetry with various functions relevant for different types
of disordered states, e.g., spin glass~\cite{Horigane2018}, inhomogeneous 
magnetic moments with short-range correlations~\cite{Dally2014,Yamauchi2015,Hallas2016,Do2018},
quantum Griffiths states~\cite{Wang2017}. Since none of them fits the data
adequately, we conclude that the degree of disorder in CePd$_{1-x}$Ni$_x$Al
is rather small compared to those states. In the Discussion section, we
elaborate in more detail about the role of disorder.

\subsubsection{Paramagnetic state}

\begin{figure}
{\centering {\includegraphics[width=8.5cm]{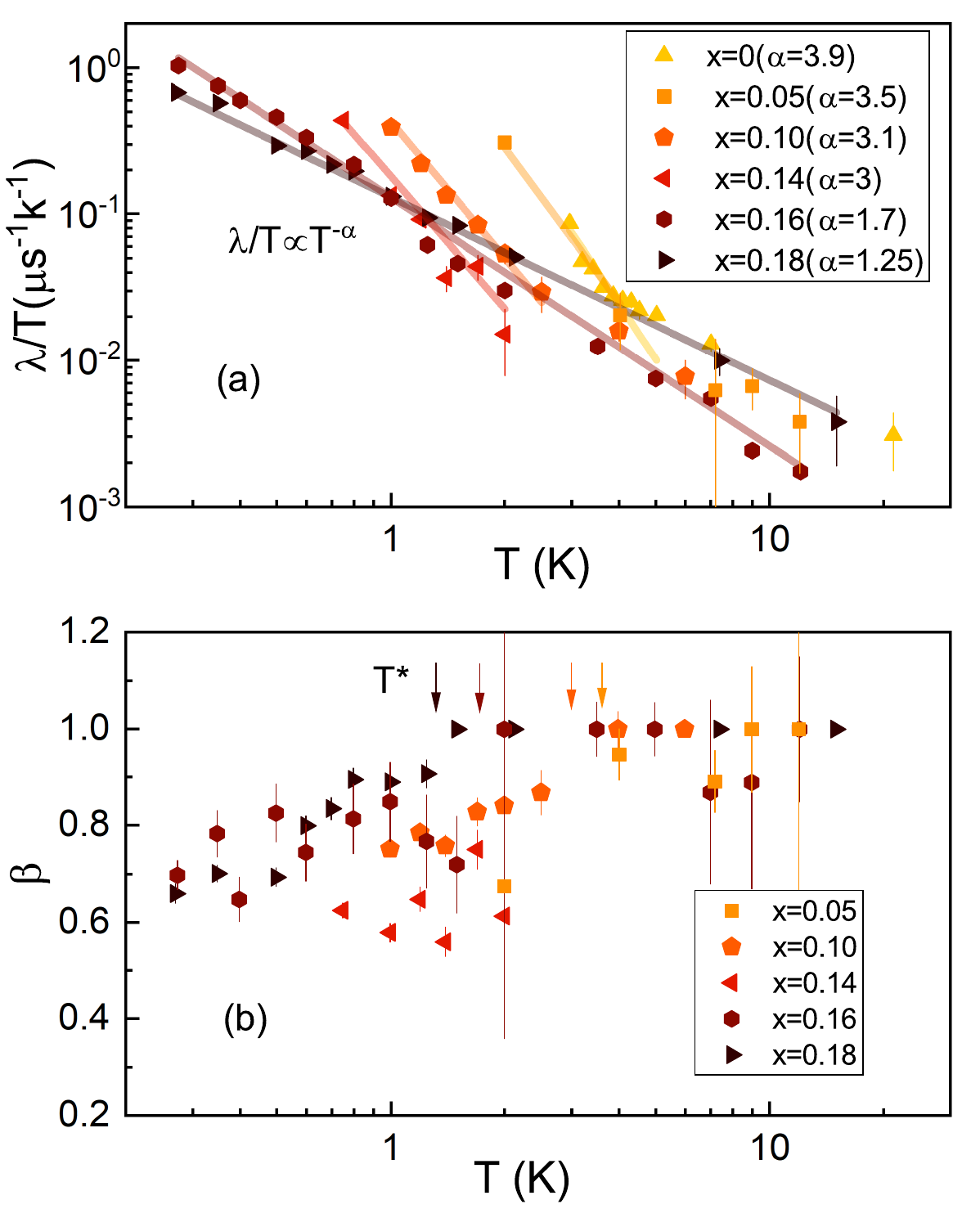}}\par} \caption{\label{fig:norm_long_relax}Temperature dependence of the longitudinal relaxation rate $\lambda/T$ for all the Ni concentrations (a). Solid lines represent power-law fits, $T^{-\alpha}$. Stretching  exponent $\beta$ for various Ni concentrations as a function of temperature (b). Arrows indicate the onset of the spin-liquid regime at $T^*$.}
\end{figure}

In the paramagnetic state, the ZF-$\mu$SR asymmetry can be well fitted
by a static Gaussian Kubo-Toyabe function, representing the nuclear moment
contribution, multiplied by a stretched exponential function, 
representing the contribution of the electronic moments: 
\begin{equation}
A(t) = A_{0}[1/3 + 2/3(1-\sigma_\mathrm{n}^2t^2)e^{-(\sigma_\mathrm{n}^2t^2/2)}e^{-(\lambda t)^{\beta}}].
\end{equation}
Here, $\lambda$ and $\sigma_\mathrm{n}$ are the exponential- and the Gaussian relaxation
rates, the latter found to be temperature independent, with
$\sigma_\mathrm{n} = 0.13$\,$\mu$s$^{-1}$. Conversely, both $\lambda$
and $\beta$ depend on temperature. At high temperatures, the value of
$\beta$ is 1, representing a fixed-field distribution and simple random
rapid fluctuations. Upon lowering the temperature, $\beta$ decreases
from 1 and $\lambda$ increases. In Fig.~\ref{fig:norm_long_relax}(a), we
show the temperature dependence of the relaxation rate
$\lambda/T=1/T_{1\mu}T \propto \chi''(q,\omega)$, where $\chi''(q,\omega)$
is the imaginary part of the dynamical spin susceptibility. For $x = 0.05$,
0.1 and 0.14, $\lambda/T$ diverges towards $T_{\rm N}$ due to the critical
slowing down of spin fluctuations by following a power-law, $T^{-\alpha}$,
whose exponent $\alpha$ decreases with increasing Ni concentration. 

More interestingly, above the critical $x_c\approx 0.15$ value
(for $x = 0.16$ and 0.18), the power-law behavior of $\lambda/T$ still
persists. Such an extension of the non-Fermi liquid regime above the
critical value of the tuning parameter (here, $x_c$) has also been seen
in pressurized CePdAl, indicating the same effect of chemical- and
hydrostatic pressure on CePdAl.

\subsection{Longitudinal field measurements}
\begin{figure*}
{\centering {\includegraphics[width=12cm]{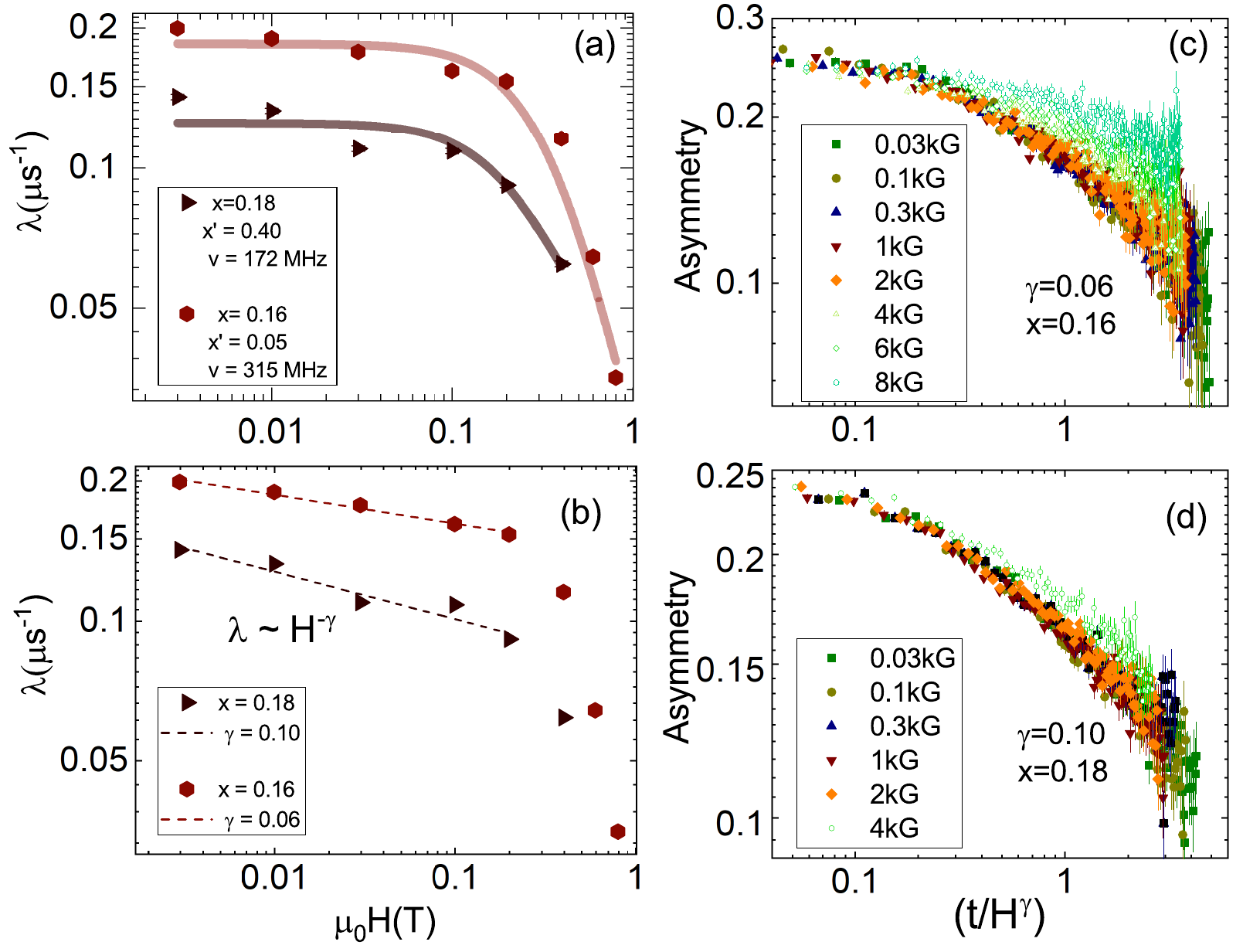}}\par} \caption{\label{fig:long_relax}$\mu$SR relaxation rate $\lambda$ vs.\ longitudinal field for $x = 0.14$, 0.16, and 0.18 (a, b). Solid lines (a) represent fits using Eq.~(\ref{eq:spin_dynamics}), while dashed lines (b) are power-law fits, $\lambda \propto H^{-\gamma}$. Time-field scaling of the $\mu$SR asymmetry at 0.27\,K for $x = 0.16$ (c) and 0.18 (d).}
\end{figure*}

$\mu$SR measurements in a longitudinal field (LF) allowed us to determine
the nature of the electron-spin dynamics close to the QCP. Figure~\ref{fig:long_relax}(a)
shows the LF dependence of $\lambda$ at 0.27\,K for $x = 0.16$ and
$x = 0.18$ (i.e., both above $x_c$). The main cause of the muon-spin
relaxation is usually the spin fluctuations of 4$f$ electrons, whose
magnetic moments couple with those of the implanted muons. To have a
quantitative idea about the nature of spin dynamics at low temperature,
we utilized the following function\cite{Zhang2022,Li2016,Cantarino2019}: 
\begin{equation}
\label{eq:spin_dynamics}
\lambda(H) = 2 \Delta^{2} \tau^{x'} \int_{0}^{\infty}\!\!\!t^{-x'} e^{-\nu t} \cos(2\pi \mu_{0} \gamma_{\mu}Ht)\,\mathrm{d}t,
\end{equation}
where $t$ is time, $\tau$ is an early-time cut-off, $\Delta$ is the width
of the internal-field distribution, $\gamma_\mu$ is the gyromagnetic ratio,
and $\nu$ is the fluctuation frequency of local moments. A fit of the
LF dependence of $\lambda$ [see Fig.~\ref{fig:long_relax}(a)] yields $x' \neq$0, which indicates that the spin-spin autocorrelation function is not a simple exponential,
$C(t) = \exp(-\nu t)$, but rather $C(t) = (\tau/t)^{x'} \exp(-\nu t)$.
The fluctuation frequency $\nu$, for $x = 0.16$ and 0.18, is estimated
to be 315\,MHz and 172\,MHz, respectively. In either case, $\nu$ is two
orders of magnitude higher than commonly found in other spin-liquid
candidate materials: NaYbS$_2$~\cite{Baenitz2018}, YbMgGaO$_4$~\cite{Li2016},
Sr$_3$CuSb$_2$O$_9$~\cite{Kundu2020}, and BaTi$_{0.5}$Mn$_{0.5}$O$_3$~\cite{Cantarino2019},
but of similar order of magnitude to other metallic heavy-fermion compounds
close to quantum criticality~\cite{Chen2022,Orain2014}. Yet, the present
$\nu$ values are slower than those of many spin-glass materials~\cite{Uemura1985}.

Interestingly, for applied magnetic fields $\mu_{0}H$ between 3 and
200\,mT, we identify also a time-field scaling $A(t/H^\gamma)$, where
$\gamma$ is an exponent. Such field dependence corresponds to evaluating
the Fourier transform of $C(t) = \langle \boldsymbol{S_i}(0) \cdot \boldsymbol{S_i}(t) \rangle$,
throughout the 0.4 to 28\,MHz frequency range ($\gamma_\mu H/2\pi$).
Time-field scaling is a signature of slow dynamics that is encountered
in both classical spin glasses and in NFL systems with local 4$f$ moments.
Independent information on the behavior of the spin-spin autocorrelation
function $C(t$) may be gained by examining the time-field scaling. While
$\gamma<1$ signifies a power-law behavior of $C(t)$, $\gamma>1$ implies
a stretched exponential behavior, typical of inhomogeneous systems. 
In Fig.~\ref{fig:long_relax}(c) and (d) we show the asymmetry vs.\ the
time-scaling variable $A(t/H^\gamma)$, for $x = 0.16$ and 0.18, respectively.
Since in both cases we find an exponent $\gamma <1$, this indicates a
power-law behavior of $C(t)$, consistent with an analysis by means of
Eq.~(\ref{eq:spin_dynamics}). The exponent $\gamma$ also matches the
power-law exponent in the $\lambda$-vs-applied-field plot shown in
Fig.~\ref{fig:long_relax}(b) up to 0.2\,T. A similar time-field scaling
has been found in UCu$_{5-x}$Pd$_x$~\cite{MacLaughlin2001},
CePtSi$_{1-x}$Ge$_x$~\cite{MacLaughlin2004}, and
CePd$_{0.15}$Rh$_{0.85}$~\cite{Adroja2008}. The high-field data
(here, above 0.2\,T) deviate from the scaling curve as well as from
the power-law behavior [Fig.~\ref{fig:long_relax}(b)], indicating that
higher applied fields have a direct effect on $C(t)$. A similar kind of
deviation has also been seen in Ce(Cu$_{1-x}$Co$_x$)$_2$Ge$_2$~\cite{Tripathi2019}.
Our LF experiments strongly suggest a cooperative spin dynamics in the
spin-liquid regime above $x_c$ and below $T^*$. \textcolor{black}{Furthermore, the LF experiments also indicate that even a field of 0.8\,T is not enough to completely suppress the relaxation process and suggests that the correlations are dynamic in nature as can be expected in the QSL state.}
%

\section{Discussion}

Our pressure-temperature phase diagram~\cite{Majumder2022} shows 
that the extended non-Fermi-liquid regime is exclusively due to
frustration, as pressure does not introduce significant disorder.
Since chemical doping may introduce disorder and an extended NFL
regime also occurs in several disordered heavy-fermion
compounds~\cite{Andrade1998, Bernal1995}, it is important to clarify
whether disorder or frustration is responsible for the magnetic
properties of CePd$_{1-x}$Ni$_x$Al above $x_{c}$.

\textit{Effect of disorder:} Disorder-induced states include the quantum
Griffiths phase (QGP), spin glasses (SG), etc. QGP appears in systems
where the quenched disorder radically affects a quantum-phase transition
giving rise to an additional QGP phase. In such a case, a smeared phase
transition can occur near the QCP. In CePd$_{1-x}$Ni$_x$Al, however,
this smearing effect is not observed (at least not down to 0.27\,K),
which is low enough with respect to the temperatures below which QGPs
have been found in other systems~\cite{Vishvakarma2021,Wang2017,Tripathi2019}.
Thus, we can discard the possibility of such a phase. It should be noted
that $\lambda/T$ obeys a power-law $T^{-\alpha}$ behavior, with
$\alpha > 1$ in the CePd$_{1-x}$Ni$_x$Al case. In other QGP systems,
however, the power-law exponent is found to be less than 1, e.g., in
Ce(Cu$_{1-x}$Co$_x$)$_2$Ge$_2$ ($\alpha = 0.55$)~\cite{Tripathi2019},
in CeCoGe$_{3-x}$Si$_x$ ($\alpha = 0.72$)~\cite{Krishnamurthy2002}, and
in CePd$_{0.15}$Rh$_{0.85}$ ($\alpha = 0.8$)~\cite{Adroja2008}.
This indicates that disorder suppresses the critical fluctuations.

Now we consider the possibility of a spin-glass (SG) state occurring in
CePd$_{1-x}$Ni$_x$Al. The absence of spin freezing in Ni-doped CePdAl
is supported by the following observations: (i) Spin glass is also a
disordered magnetic state, where spins are statically aligned randomly,
so that the dynamics is very slow, usually between 0.01 and 0.1\,ns~\cite{Uemura1985}.
Our LF-$\mu$SR measurements indicate a time scale of about 3--5\,ns,
more than an order of magnitude larger than that of SG systems (mentioned in
the previous section). (ii) The stretching exponent $\beta$ decreases gradually from 1 at high temperatures to a constant value of 0.6 at low temperatures [see Fig.~\ref{fig:norm_long_relax}(b)]. On the other hand, for spin-glass systems, $\beta$ is expected to decrease to 1/3 at the spin freezing temperature~\cite{Keren1996,Campbell1994}. (iii) As already discussed in the previous section, we were unable to fit the $\mu$SR asymmetry by considering any function relevant for the spin-glass state.

At the same time, resistivity measurements at higher Ni doping~\cite{Fritsch2014}
show a negligible distribution of Kondo temperatures. This further
indicates that Ni doping introduces at most only a weak disorder.
Thus, it can be stated that the weak disorder introduced by Ni doping,
is unable to form neither QGP nor SG states in CePd$_{1-x}$Ni$_x$Al. 
The quasi-static state observed for $x = 0.1$ and 0.14 may be an
unconventional, weak disorder-induced state. Along with the above
mentioned points, the incompatibility of fitting the asymmetry with
different fit functions used for various disordered states
supports the absence of strong disorder effects. 

Usually, the lower the value of the exponent $\gamma$, the lesser the chance of having a disordered ground state. Thus, in the present case, $\gamma \ll 1$ (discussed in the previous section) excludes a
dominant effect of disorder in the critical fluctuations around the QCP.
Since the effect of chemical- and hydrostatic pressure on
CePdAl is identical \textcolor{black}{(as far as the magnitude of $\lambda/T$ and the
power-law exponents are concerned),} 
this again supports the weak effect of disorder in determining the
non-Fermi liquid regime in this compound.

\textit{Effect of frustration:} It is interesting to note that, for
$x=0$, 0.05, and 0.1, the enhancement of $\lambda/T$ [see Fig.~\ref{fig:norm_long_relax}(a)]
starts from a temperature tenfold above $T_{\rm N}$, indicating a
high degree of frustration in CePd$_{1-x}$Ni$_x$Al. Neutron scattering
also detects short-range correlations at a temperature much higher than
$T_{\rm N}$. Also for $x = 0.16$ and 0.18 the enhancement of $\lambda/T$
from a very high temperature suggests the significant role of frustration
on the quantum critical nature of spin fluctuations. Note that, such an
enhancement of $\lambda/T$ from a very high temperature,
was not observed in other Ce-based systems~\cite{Tripathi2022,Tripathi2019,Adroja2008}
lacking a geometrically frustrated lattice. 
This implies that frustration persists throughout the full doping range
studied in the present work and it provides evidence that frustration
is responsible for the extended non-Fermi-liquid regime (at $x > x_{c}$)
in CePdAl. 

The presence of finite frustration is also reflected in the temperature
dependence of $\beta$ shown in Fig.~\ref{fig:norm_long_relax}(b). Our
pressure dependent $\mu$SR study identifies a characteristic temperature  
$T^*$, below which $\beta$ deviates from 1. This was assigned to the
temperature below which spin-liquid-like correlations develop~\cite{Majumder2022}.
In the present case, too, such a characteristic temperature $T^*$ exists
and is found to decrease with increasing Ni concentration, as shown in
Fig.~\ref{fig:norm_long_relax}(b). This again confirms the similarity
between the hydrostatic- and chemical-pressure effects in the magnetic
properties of Ce\-Pd\-Al. These observations indicate the importance of
frustration on the quantum critical fluctuations and also the persistence
of frustration throughout the full doping range. 

From the above discussion, we have established that disorder has only
a limited role in originating the non-Fermi liquid regime in
CePd$_{1-x}$Ni$_x$Al, while the persistent frustration is clearly
responsible for the extended NFL regime. The weak disorder seems to
affect and stabilize only the quasi-static ordered state for $x < x_{c}$.  

\section{conclusion}
In conclusion, by utilizing the local-probe $\mu$SR technique, we investigated the quantum
critical Ni-doped CePdAl system across the full range of Ni concentrations,
both above and below $x_{c} = 0.15$. 
For $x = 0.05$, we observe an incommensurate LRO state, similar to that
of the CePdAl parent compound. Upon increasing the Ni concentration,
disorder appears and the LRO changes into a quasi-static magnetic
order for $x = 0.1$ and 0.14. This state does not exhibit the properties
expected for a conventional spin-glass- or a quantum Griffiths state,
thus indicating a relatively low amount of Ni-induced disorder in this system. 
Interestingly, a power-law dependence of the longitudinal relaxation rate
and a time-field scaling are observed for $x = 0.16$ and 0.18, suggesting
a non-Fermi liquid regime. Since both values lie above $x_{c}$, our
results indicate that the non-Fermi liquid regime extends beyond $x_c$,
thus mimicking the behavior of the pressure-induced phase diagram
of CePdAl. Since frustration is shown to persist up to $x = 0.18$,
the extended non-Fermi liquid regime is, therefore, due to frustration.
Akin to the pressure effect on the magnetic properties of pure CePdAl,
the chemical pressure in the Ni-doped case also reduces a characteristic
temperature $T^*$, below which spin-liquid-like dynamical fluctuations
occur in the $x = 0.16$ and 0.18 case. In summary, all our observations
suggest a similar role for the hydrostatic- and chemical pressure on
the magnetic properties of CePdAl, with frustration being the main
driving force behind such unusual behavior. In the future it might be interesting to see how far (in terms of Ni doping) such a frustration-driven non-Fermi liquid regime persists until the Fermi liquid emerges.

\section{Author contributions}
V.F. and O.S. prepared the sample and initiated the study of CePdAl. T.S. performed the $\mu$SR measurements, with the online assistance of I.I and M.M.. I.I and M.M. analysed the data. I.I., M.M. and T.S. wrote the manuscript with notable inputs from all the co-authors. M.M. supervised the project.

\end{document}